\documentclass[aps,twocolumn,showpacs,superscriptaddress,groupedaddress, nofootinbib]{revtex4-1}

\usepackage{hyperref}
\usepackage{multirow}
\usepackage{mathrsfs,amsmath} 
\usepackage{verbatim}
\usepackage{amsmath}
\usepackage{amsfonts}
\usepackage{amssymb}
\usepackage{amsthm}
\usepackage{bm}
\usepackage{xparse}
\usepackage{xcolor}
\usepackage{textcase}
\usepackage{url}
\usepackage{lipsum}
\usepackage{appendix}
\usepackage{dsfont}
\usepackage{epstopdf}
\usepackage{footnote}
\usepackage{tgbonum}
\usepackage{blindtext}
\usepackage{enumitem}
\usepackage{mathrsfs,amsmath} 
\usepackage{float}
\usepackage{subfloat}
\usepackage{wrapfig}
\usepackage{pslatex}
\usepackage{amsmath}
\usepackage{amsfonts}
\usepackage{setspace}
\usepackage{epstopdf}
\usepackage{wrapfig}
\usepackage{csquotes}
\usepackage{hyperref}
\usepackage{graphicx}
\usepackage{amssymb}
\usepackage{multirow}
\usepackage{xcolor}
\usepackage{framed}
\usepackage{mathtools}
\usepackage{tcolorbox}

\definecolor{shadecolor}{rgb}{0.8,0.9,1}
\DeclareDocumentCommand{\Tr}{m m O{\big}}{{\rm Tr}_{\:\!{#1}}#3({#2}#3)}

\begin{document}
\title{Reply to a ``Comment on `Physics without determinism: Alternative interpretations of classical physics' ''}

\author{Flavio Del Santo}
\affiliation{
Institute for Quantum Optics and Quantum Information (IQOQI-Vienna), A-1090 Vienna, Austria; and
Faculty of Physics, University of Vienna, A-1090 Vienna, Austria}

\author{Nicolas Gisin}
\affiliation{Group of Applied Physics, University of Geneva, 1211 Geneva 4, Switzerland}

\date{\today}

\begin{abstract}
In this short note we reply to a comment by Callegaro \textit{et al.} \cite{callegaro}  that points out some weakness of the model of indeterministic physics that we proposed in Ref. \cite{delsantogisin}, based on what we named ``finite information quantities'' (FIQs). While we acknowledge the merit of their criticism, we maintain that it applies only to a concrete example that we discussed in \cite{delsantogisin}, whereas the main concept of FIQ remains valid and suitable for describing indeterministic physical models. We hint at a more sophisticated way to define FIQs which, taking inspiration from intuitionistic mathematics,  would allow to overcome the criticisms in \cite{callegaro}.
\end{abstract}

\maketitle

In this short note we reply to a “Comment on ‘Physics without determinism: Alternative interpretations of classical physics'” \cite{callegaro} by Callegaro \textit{et al.} that points out some limits of the model of indeterministic physics that we proposed in Ref. \cite{delsantogisin}. In particular, their criticism demonstrates a weakness in the definition of finite information quantities (FIQs), which were introduced in that paper to address the fundamental problem that real numbers --assumed to be the values taken by physical variables-- contain in general an infinite amount of information.

In Ref. \cite{delsantogisin}, we defined a FIQ as a quantity  $Q \in [0,1]$ expressed in binary basis, $Q=0.Q_1Q_2Q_3…$, where to each of its bits $Q_k$ is associated a “propensity” $q_k$, i.e. an objective property that quantifies the (possibly unbalanced) disposition or tendency of the $k$-th bit to take the value 1 (as opposed to its complementary value 0, which would have an associated propensity of $1-q_k$). In this way, physical quantities are conceived as indeterministic and it is the vector of the propensities associated to each bit $[q_1, q_2, …q_k …]$ that  completely characterizes a physical quantity (i.e. it represents a ``pure state'' for that quantity). Note, however, that by construction propensities ought to satisfy certain constraints on the total information content, in order for a quantity to be a FIQ. Incidentally, it should be noticed that –contrarily to what the authors of the comment \cite{delsantogisin} maintain– the concept of propensities as expressed here may not be exactly formally equivalent to that of probability. In fact, it was postulated that they take value only in the rational numbers and they may even not satisfy all Kolmogorov’s axioms (as showed, e.g., in Ref. \cite{gisin91}). 

Having introduced the vector of propensities, a FIQ is then defined through the following necessary condition: It is necessary for a vector of propensities that its information content (as expressed by some reasonable measure) is finite. We still believe that this minimal necessary requirement represents a good definition of a finite information quantity, which allows an indeterministic view of (classical) physics.
Yet, in \cite{delsantogisin}, we also expressed a sufficient condition for a quantity to be a FIQ, namely that there exist a value $M$, after which all the bits are equally likely to take value 0 or 1, i.e. such that its vector of propensity takes the form $[q_1, q_2, …q_k ,… q_M, \frac{1}{2}, \frac{1}{2},…]$. The fulfillment of this latter condition implies that the bits of a FIQ are mutually independent, and the authors of the comment \cite{callegaro} successfully showed that this is a weakness of our proposal. Indeed, through the introduction of a minimal arithmetic for FIQs, they demonstrated that, under a basic operation such as a change of units (which in practice is obtained by multiplying a FIQ by a constant number), the mutual independence of the bits of a FIQ is not preserved. We were aware that our sufficient condition for a physical quantity to be a FIQ was based on the implicit assumption of the independence between bits and that this could lead to problematic issues (e.g., when changing units or even the numerical base). Hence, we acknowledge the merit of Ref. \cite{callegaro} in formally showing why it is so.

It thus seems to us that this criticism only applies to the sufficient condition, which was primarily introduced as a concrete and intuitive example of how a FIQ could be constructed. In fact, this issue could be overcome by resorting to more sophisticated examples that introduce correlations between the bits, while still fulfilling the constraints on the finiteness of information as imposed by the necessary condition above. This is totally admissible within the original definition of FIQs, as long as the necessary condition of finite information content of the vector of probability is met. Since in Ref. \cite {delsantogisin} we only considered independent bits, we adopted,  as a measure of the information content, the (infinite) summation of the complementary to the binary entropy of the propensity of each bit. If we now are to consider, also in the light of the criticism in \cite{callegaro}, correlated bits, a different measure of information content is required, such as a more complex function of the binary entropy. 

One possible way to refine the definition of FIQs such that it fulfills the necessary condition but does not incur in the criticism raised in \cite{callegaro}, is to take inspiration from intuitionistic mathematics, as one of us (N.G.) is currently developing (see \cite{gisin2, gisin3}). In a nutshell, assume that nature has the power to continually produce genuinely new information in the form of a random bit $r(n)$ produced at every discrete time instants $n$. Let $k$ be an odd positive integer and define the $n$th bit of a FIQ by the majority vote of the $k$ last random bits $r(n-k+1)$,..,$r(n)$. In this way, at each time instant $n$, the FIQ is determined by a finite amount of information, and yet the bits of the FIQ are correlated. Two such basic FIQs can be added and multiplied (like in the case of a change of units as proposed in the comment) using the usual arithmetic rules resulting in new finite information quantities. This procedure extends to all standard arithmetic.

Certainly, while resolving the issues raised in \cite{callegaro} with the independent bits, some questions of fundamental nature remain open in this new definition of FIQs, such as, e.g., the characterization of the odd integer $k$ which may depend on the specific dynamical system under investigation. However, FIQs remain a promising conceptual tool to formalize our physical world: A world that does not need to attribute a physical reality to abstract mathematical entities that contain infinite information and, therefore, where not everything is bound to happen with certainty.

\begin{small}

\end{small}

\end{document}